\documentclass[referee]{raa}            % referee version: for submission

\usepackage{graphicx,times}             %for PS/EPS graphics inclusion, new
\usepackage{natbib}
\usepackage{amssymb,amsmath}

\begin{document}

   \title{ Probing the large-scale structure of the universe through gravitational wave observations
%\footnotetext{$*$ Supported by the National Natural Science Foundation of China.}
}
%   \subtitle{I. Place Your Subtitle Here}

   \volnopage{Vol.0 (20xx) No.0, 000--000}      %%preserved for Editor. DOn't remove!
   \setcounter{page}{1}          %%starting page, preserved for Editor. DOn't remove!

   \author{Xiaoyun Shao
      \inst{1}
   \and Zhoujian Cao
      \inst{1}
   \and Xilong Fan
      \inst{2}   
   \and Shichao Wu
      \inst{3,4}
   }

   \institute{Department of astronomy, Beijing Normal University 100875, China\\
        \and
            School of Physics and Technology, Wuhan University, Wuhan, Hubei 430072, China; {\it xilong.fan@whu.edu.cn}\\
        \and
            Max-Planck-Institut f{\"u}r Gravitationsphysik (Albert-Einstein-Institut), D-30167 Hannover, Germany \\
        \and
            Leibniz Universit{\"a}t Hannover, D-30167 Hannover, Germany \\
\vs\no
   {\small Received~~20xx month day; accepted~~20xx~~month day}}
   
\abstract{{The improvements in the sensitivity of the gravitational wave (GW) network enable the detection of several large redshift GW sources by third-generation GW detectors. These advancements provide an independent method to probe the large-scale  structure of the universe by using the clustering of the binary black holes. The black hole catalogs are complementary to the galaxy catalogs because of large redshifts of GW events, which may
imply that binary black holes (BBHs) are a better choice than galaxies to probe the large-scale structure of the universe and cosmic evolution over a large redshift range. To probe the large-scale structure, we used the sky position of the binary black holes observed by third-generation GW detectors to calculate the angular correlation function (ACF) and the bias factor of the population of binary black holes. This method is also statistically significant as 5000 BBHs are simulated. Moreover, for the third-generation GW detectors, we found that the bias factor can be recovered to within 33$\%$ with an observational time of ten years. 
This method only depends on the GW source-location posteriors; hence, it can be an independent method to reveal the formation mechanisms and origin of the BBH mergers compared to the electromagnetic
method.}
}

   \authorrunning{XiaoYun Shao, et al. }            %author_head in even pages
   \titlerunning{Probe the large-scale structure by GW}  % title_head in odd pages

   \maketitle
%% The author head (on even pages) 
\section{Introduction}           
\label{sect:intro}
{ Gravitational waves (GWs) were successfully detected by the Laser Interferometer Gravitational Wave Observatory (LIGO) on September 15, 2015 (\citealt{6}). There have been several black hole binaries detected
by the Advanced LIGO and Virgo interferometers (\citealt{c1}; \citealt{c2}; \citealt{c4}; \citealt{c5}; \citealt{c6}; \citealt{c7}; \citealt{c8}; \citealt{c9}; \citealt{c10}). The First GW Transient Catalog (GWTC-1) from the first two observing runs reported 11 GW events from compact binary mergers (\citealt{c6}), which was published by the LIGO-Virgo Collaboration. The Second GW Transient Catalog (GWTC-2) includes a total of 44 confident
binary black hole events (\citealt{c11}; \citealt{c12}). With the increasing number of detected events, the observation of GWs has heralded a new era.

Based on the LIGO-Virgo detectors, most GW signals originate from the merger of compact objects (\citealt{c2}). In coming years, KAGRA (\citealt{r9}) and LIGO-India (\citealt{r10}) could join a network of GW observatories. The proposals of the third-generation detectors (3G) greatly improve the sensitivity of a network of detectors. The proposed 3G detectors include LIGO Voyager (\citealt{r12}; \citealt{r13}), Einstein Telescope (ET) (\citealt{r12}; \citealt{r14}), and Cosmic Explorer (CE) (\citealt{r12}; \citealt{r15}).

GWs could become a new way to probe cosmology in the future. The Hubble constant is measured following the multi-messenger detection of the binary neutron star merger GW170817 with measured value $H_{0} = 70.0_{-8.0}^{+12.0}~$km$~$s$^{-1}~$Mpc$^{-1}$ (\citealt{13}). By using the multi-messenger method, 50 detections of binary neutron star mergers by the Advanced LIGO and Virgo interferometers may lead to a precise measurement of the Hubble constant (\citealt{p9}; \citealt{p10}) and resolve its inconsistent measurements (\citealt{schutz1986determining}; \citealt{13}).

%\textbf{\color{red}{With hundreds of thousand or millions of GW events detected, we can use their mergers as tracers of Large Scale Structures over a large redshift range (\citealt{14}; \citealt{15}; \citealt{16}; \citealt{17}; \citealt{18}). The tracer with binary black holes is independent and complementary to the tracer with galaxies clustering. Because of large redshifts of GW events (z is up to 10) (\citealt{ligo2019}), the black holes catalogs will be complementary to the galaxy catalogs, which may suggest binary black holes are better than galaxies to probe the large-scale distribution and cosmic evolution  over a large redshift range. Beyond this aspect, we can characterize their clustering properties and trace Dark Matter distribution with cosmological bias by the statistical studies of the spatial distribution of GW events (\citealt{a58}). The studies that the spatial distribution of binary black hole merger are anisotropically distributed are carried on (\citealt{a49}). GW bias providing information on the physical nature of the different mergers can be meassured by cross-correlation between galaxy and GW surveys (\citealt{17}; \citealt{e52}). So we can know whether black hole mergers trace more closely the distribution of dark matter which suggests the primordial origin of BBHs or that of stars harboured in luminous and massive galaxies which suggests a stellar origin of BBHs (\citealt{e53}). Another approach to measuring GW bias based on source-location posteriors and two point correlation function is proposed (\citealt{23}). }}%

We can use the millions of BBHs observed by future GW detector as tracers of large-scale structures over a large redshift range. The large-scale structure is studied by using the number density and luminosity distances observed by future GW detectors (\citealt{15}; \citealt{18}). The studies find that the distribution of BBHs mergers within GWTC-1 is isotropic and the anisotropy can be probed with an angular scale of degrees at the level of $<=0.1\%$ within an observational time of a year (\citealt{a49}). The tracer with binary black holes is independent and complementary to the tracer with galaxy clustering. The black hole catalogs are complementary to the galaxy catalogs because of large redshifts of GW events (z is up to 10) (\citealt{ligo2019}), which may suggest that binary black holes are a better choice than galaxies to probe the large-scale structure of the universe and cosmic evolution over a large redshift range. Beyond this aspect, we can characterize their clustering properties and trace dark matter distribution with cosmological bias by statistically studying the spatial distribution of GW events. The matter inhomogeneities in the universe can be traced by the spatial distribution of BBHs. Its results show that the clustering bias of BBHs observed by the second-generation GW detectors are greater than or equal to 1.5 with an observational time of three years (\citealt{16}). GW bias, which provides information on the physical nature of the different mergers, can be measured by cross-correlation between the galaxy and GW surveys (\citealt{17}; \citealt{e52}; \citealt{a58}). Therefore, we can determine whether black hole mergers trace the distribution of dark matter more closely than other traditional methods, which may suggest a primordial origin of BBHs, or that of stars harbored in luminous and massive galaxies, leading to a stellar origin of BBHs (\citealt{e53}). Another approach to measuring GW bias based on source-location posteriors and two-point correlation function is proposed (\citealt{23}).}

Although statistics can be derived from merging events of double black holes for constructing the large-scale structure of the universe based on GW detection (\citealt{14}; \citealt{15}; \citealt{16}; \citealt{17}; \citealt{18}), the positioning accuracy of GW source measured by a ground-based detector is poor. The error in the angular direction is more than 10 square degrees and that of the distance is more than 10$\%$ (\citealt{19}). Even if an ideal space detector network is used, the error in the angular direction could exceed 1 square arcsec, and the error of the distance could exceed 10 $\%$ (\citealt{20}; \citealt{21}; \citealt{22}). Therefore, the feasibility to construct the large-scale structures of the universe by GW detection is questioned.

{ Recently, a research group neglected the correlations between the errors in right ascension, declination, and distance. Based on this assumption, the team concluded that the third-generation GW detectors can build large-scale structures of the universe as long as the error in the angular direction is less than 1 square degree and the error of distance is less than $\sim 90~h^{-1}$ Mpc (\citealt{23}). We rejected the above assumption to calculate more accurate observations. We assume that the position measurement errors in the three directions of space are correlated, and the value of the error depends on the distance of the GW source in our present study. We use the angular correlation function to explore the effect of estimating the bias factor from the spatial distribution of the observed BBH population, which is useful in identifying the GW events' origins.}

This paper is organized as follows: We describe how to theoretically calculate the smeared angular correlation function, where the relationship between the angular correlation function and the smearing angular correlation function is derived in Section \ref{sec:method}. We detail the simulation of the BBH catalogs and estimate the bias factor by using the least square method in Section \ref{im}. We discuss the results in Section \ref{sec:sum}.

$~~~~~~~~~~~~~~~~~~~~~~~~~~~~~~~~~~~~~~~~~~~~~~~~~~~~~~~~~~~~~~~~~~~~~~~~~~~~~~~~~~~~~~~~~~~~~~~~~~~~~~~~~~~~~~~~~~~~~~~~~~~~~~~~~~~~~~~~~~~~~~~~~~~~~~~~~~~~~~~~~~~~~~~~~~~~~~~~~~~~~~~~~~~~~~~~~~~~~~~~~~~~~~~~~~~~~~~~~~~~~~~~~~~~~~~~~~~~~~~~~~~~~~~~~~~~~~~~~~~~~~~~~~~~~~~~~~~~~~~~~~~~~~~~~~~~~~~~~~~~~~~~~~~~~~~~~~~~~~~~~~~~~~~~~~~~~~~~~~~~~~~~~~~~~~~~~~~~~~~~$

%%This paper is organized as follows. In Section \ref{sec:method}, we describe how to calculate smeared angular correlation function theoretically. And we details how to simulate the BBH catalogs and estimate the bias factor in Section \ref{im}. We show and discuss the results in Section \ref{sec:sum}.
%This paper is organized as follows. The relationship between angular correlation function and two point correlation function is described in Section \ref{subsec:acf}. In Section \ref{subsec:estacf}, we describe how to estimate angular correlation function using Landy-Szalay (LS) estimator (\citealt{24}). In Section \ref{subsec:distance}, we describe  measurement error of BBHs' location in the simplified assumptions. The relationship between the angular correlation function and the smearing angular correlation function is derived in Section \ref{subsec:thero}. In Section \ref{c0}, we detail GW events’ error matrixes in our work. And in Section \ref{subsec:sim}, we describe how to simulate the mock BBH catalogs, estimate the bias factor by using least square. We show and discuss the results in Section \ref{sec:sum}.
%% Authors can give a citation as 'Michel et al. 1992'.
%% You may also use \cite, \citep and \citet for citation, and use Table~1 or Figure~1
%% and so forth. Using \ref and \label for cross-references of Tables/Figures
%% is a good way in adjusting/adding/removing text, tables or figures.

\section{Theory}
\label{sec:method}
\subsection{Angular correlation function (ACF)}
\label{subsec:acf}
The auto-correlation function (ACF) $\omega(\theta)$ is defined as the excess probability in comparison with a random Poisson distribution, which can be expressed as a function of angular separation $\theta$ on the sky. We can express the ACF through the two-point correlation function $\xi(R)$ using the Limber approximation (\citealt{sim})
%We can simply convert the two point correlation function $\xi(R)$ to the angular correlation function $\omega(\boldsymbol{\theta})$ using the Limber approximation (\citealt{sim}), and is given by
\begin{equation}
\label{1}
w(\theta) \approx \int_{0}^{\infty} d r_{1} \int_{0}^{\infty} d r_{2} p_{1}\left(r_{1}\right) p_{2}\left(r_{2}\right) \xi(R).
\end{equation}
where 
\begin{equation}
R \equiv \sqrt{r_{1}^{2}+r_{2}^{2}-2 r_{1} r_{2} \cos \theta.}
\end{equation}
and
\begin{equation}
  P_{1}=P_{2}=P(r) \left\lbrace \begin{array}{ll}~0 ~~~~~~0\leq r <r_{1}\\
                                              \frac{1}{r_{2}-r_{1}} ~~~~~~r_{1}\leq r < r_{2} \\
                                              ~0 ~~~~~~r\geq r_{2},
                             \end{array} \right.
\label{eq:Harr}
\end{equation}
where $r_{1}$ and $r_{2}$ are the distances of two GW events from the observer, and $R$ is the distance separation between the two events.

{ The observed BBH population can be used to reveal traces of dark matter because dark matter is more abundantly found than light-emitting matter (\citealt{pla}). }The ability to trace dark matter can be quantitatively expressed as (\citealt{23}) 
\begin{equation}\label{ksi1}
\xi_{\mathrm{BBH}}(r)=b_{\mathrm{BBH}}^{2} \xi_{\mathrm{DM}}(r).
\end{equation}
With the absence of measurement errors, we can then use the ACF of dark matter and the bias factor of BBH to represent the ACF of BBHs, which is derived by using the Equation \ref{1} and \ref{ksi1}, and is given by
%, we can write (in the absence of any measurement errors)
\begin{equation}\label{b}
\omega_{BBH}(\theta)=b_{BBH}^{2} \omega_{DM}(\theta).
\end{equation}

\subsection{Estimating the ACF} \label{subsec:estacf}

\begin{figure}
   \centering
   \includegraphics[width=\textwidth, angle=0]{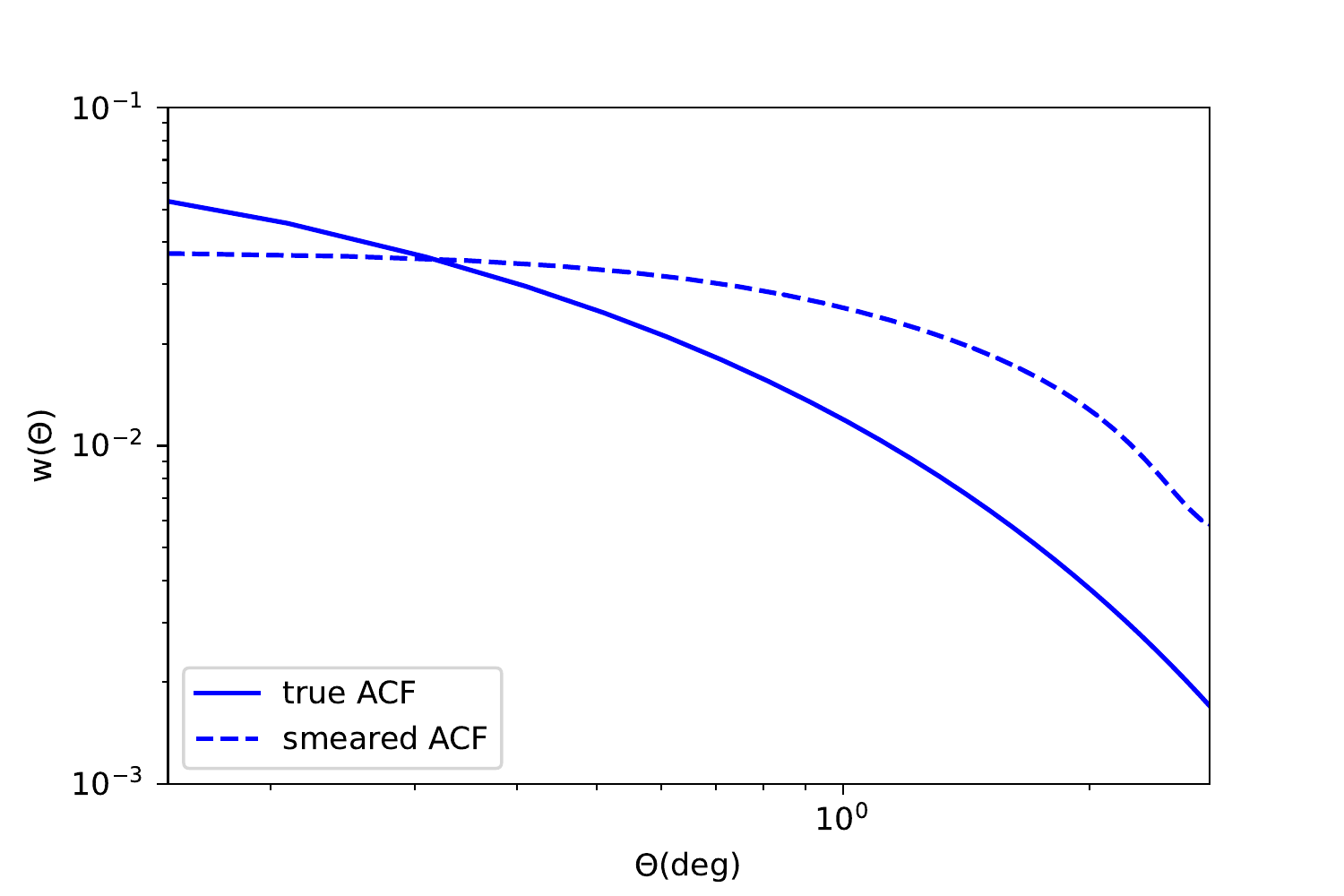}
   \caption{The results of the smeared ACF (dashed line) and the true ACF (solid) at redshift 0.3 are plotted. 
   The former assumes that the source location errors follow a Gaussian Distribution with the covariance matrix proportional to the square of the distance from the GW source. 
   The latter is calculated by \citet{36} in standard cosmology model. }
   \label{general}
   \end{figure}

The Landy-Szalay (LS) estimator can estimate the ACF from the data of the population of points according to the definition of $\omega(\theta)$, { that is, the excess probability of seeking out two points with an interval of $\theta$ as compared to a random Poisson distribution, and is given by
\begin{equation}
\omega(\theta)=\frac{D D(\theta)-2 D R(\theta)+R R(\theta)}{R R(\theta)},
\label{ls}
\end{equation}
where $DD(\theta)$, $DR(\theta)$, and $RR(\theta)$ are the number of data-data, data-random and random-random pairs normalized by the number of random-random pairs in each angular bin, respectively, (\citealt{24}). 

Several BBH mergers could be observed by using ET and CE detectors; thus, the clustering of BBHs could be explored by estimating $\omega(\theta)$ with the BBHs spatial distribution. However, the precision in the GW source location could be poor as compared to the galaxies location (which can be described as a point in the survey volume). Therefore, the observed ACF of BBHs must be different from the true ACF because of the large measurement error of the GW source location, implying that the poor source localization distributes weights from the points of the actual location to a smeared field around those points. The distribution of the GW source localization uncertainties from the population affects the smeared ACF. By convolving ACF with the GW source-location posteriors, the smeared ACF can be computed (Figure~\ref{general}).}
%(which is described as a point in the survey volume)%. 

This study assumes that the two-dimensional source-location posterior distribution of the GW event satisfies the Gaussian distribution with its true position as the mean value, and the error matrix is proportional to the square of the distance from the GW source. Future GW detectors with high sensitivity can perform reasonably well because the measurement error is inversely proportional to the square of the signal-to-noise ratio (SNR), and the SNR is inversely proportional to the distance from the GW source (\citealt{1993PhRvD..47.2198F}). A normalized
total posterior distribution in right ascension and declination $P(\mathbf{\boldsymbol{p}})=\frac{1}{N} \sum_{i=1}^{N} P_{l}(\mathbf{\boldsymbol{p}})$ can be denoted by combining $N$ simulated GW source-location posterior distributions, and is given by
\begin{equation}
P_{l}(\mathbf{\boldsymbol{p}})=\mathcal{N} \exp \left[-\frac{1}{2}\left(\mathbf{\boldsymbol{p}}-\boldsymbol{m}_{l}-\Delta \boldsymbol{m}_{l}\right)^{T} \mathrm{C}_{l}^{-1}\left(\mathbf{\boldsymbol{p}}-\boldsymbol{m}_{l}-\Delta \boldsymbol{m}_{l}\right)\right],
\label{p}
\end{equation}
where $\mathbf{\boldsymbol{p}}=\{\alpha,~\beta\}$, $\alpha$ is right ascension, $\beta$ is declination, and $\boldsymbol{m}_{l}$ is the true location in two dimensions of $l^{th}$ BBHs. The covariance matrix of the Gaussian distribution of $l^{th}$ GW source location is denoted as $C_{l}$, while $\mathcal{N}=1 / \sqrt{(2 \pi)^{2}\left|\mathrm{C}_{l}\right|}$ is a normalized factor related to distance. 
The detector noise introduces the scatter $\Delta \boldsymbol{m}_{l}$ so that the individual GW event posterior is not centered on the true location.
$\Delta \boldsymbol{m}_{l}$ satisfies the Gaussian distribution of the zero-mean and the error matrix $C_{l}$.%Figure~\ref{fig:2} shows the P(x) from a simulated catalog of BBH observations.%

The smeared ACF can then be calculated via LS estimator after sampling from the Gaussian distribution. After repeating these steps many times (see Section \ref{subsec:sim}), we can get the average value and error bars of smeared $\omega_{BBH}$. The software package Corrfunc (\citealt{39}) is used to estimate the smeared black hole ACF. It is contrasted with the smeared black hole ACF deduced by the theoretical dark matter correlation function and bias factor (see Section \ref{subsec:thero}), so that the bias factor $b_{BBH}$ can be recovered.

\subsection{The distance effect } 
\label{subsec:distance}
For more accurate observational GW source-location posteriors detected by future GW detectors, we assume that the error matrix of the GW source-location posterior is proportional to the square of the distance from the GW source as the measurement error is inversely proportional to the square of the SNR, and the SNR is inversely proportional to the distance from the GW source. The correlations between the errors in right ascension $\alpha$ and declination $\beta$ are also considered.

In this study, GW source-location posteriors are assumed to be a Gaussian Distribution with a covariance matrix $C_{l}$, which is given by
\begin{eqnarray}
P_{l}\left(\mathbf{\boldsymbol{p}}-\boldsymbol{m}_{l},~ \Delta \boldsymbol{m}_{l},~d_{l}\right)=\frac{1}{\sqrt{(2 \pi)^{2}\left|\mathrm{C}_{l}\right|}} \exp \left[-\frac{1}{2}\left(\mathbf{\boldsymbol{p}}-\boldsymbol{m}_{l}-\Delta \boldsymbol{m}_{l}\right)^{T}\right.  \left.\times \mathrm{C}_{l}^{-1}\left(\mathbf{\boldsymbol{p}}-\boldsymbol{m}_{l}-\Delta \boldsymbol{m}_{l}\right)\right].
\label{8}
\end{eqnarray}
The covariance matrix $C_{l}$ depends on the SNR and source sky location ($\alpha$,~$\beta$).
Next, we deduce the relationship between the covariance matrix of the GW source-location posterior and the distance from the GW source. For simplicity, we only deduce it in the situation when the covariance matrix is diagonal; however, our conclusion remains valid when the covariance matrix is non-diagonal. The diagonal matrix is written as,
\begin{eqnarray}
C= \left(
\begin{array}{cccc}
\sigma_{1}^{2} & & 0 \\
0 & & \sigma_{2}^{2} 
\end{array}     \right), 
\end{eqnarray}
where $\sigma_{1}$ and $\sigma_{2}$ are respectively 1-sigma error of the posterior distributions of sky location in right ascension and declination. The errors in the location $\Delta \Omega$ can be written as,
\begin{equation}
\Delta \Omega \propto \sqrt{\sigma_{1}^{2}\sigma_{2}^{2}}.
\end{equation}
\begin{equation}
\Delta \Omega \propto \frac{1}{\rho^{2}}.
\label{rou}
\end{equation}
Here we ignore the sky location dependence and relate the SNR to the distance:
\begin{equation}
\rho^{2} \propto \frac{1}{d^{2}},
\end{equation}
where $\rho$ is the SNR, and $d$ is the distance from the GW source.
Therefore we can derive the following:  
%\begin{equation}\Delta \Omega \propto d^{2}.\label{cd}\end{equation}%%
\begin{equation}
{\sigma_{1}}^2 \propto d^{2}.
\end{equation}.
\begin{equation}
{\sigma_{2}}^2 \propto d^{2}.
\end{equation}
Then we can verify that the covariance matrix of the GW source-location posterior is proportional to the square of the distance from the GW source, which is written as 
\begin{equation}
C \propto d^{2}.
\end{equation}
Thus, the covariance matrix for the individual GW source is given by: 
\begin{equation}
C_{l}=a* \left(\frac{d_{l}}{d_0}\right)^2*C_{0},  
\label{ci}
\end{equation}
where $C_0$ is a reference covariance matrix for the GW source at the distance $d_0$, and $a$ is scaled by the sensitivity improvements in the detectors.

\subsection{The smeared theoretical black hole ACF} \label{subsec:thero}

We calculate the true ACF of BBH mergers before deducing the smeared ACF. The total source-location posterior in the two dimensions of BBH mergers with no measurement error is denoted as  
\begin{equation}
P_{\mathrm{0}}(\boldsymbol{m})=\frac{1}{N} \sum_{l} \delta^{(2)}\left(\boldsymbol{m}-\boldsymbol{m}_{l}\right),
\label{8}
\end{equation}
where $\delta^{(2)}$ and $\boldsymbol{m}_{l}$, respectively, denote the two dimensional Dirac delta function and the sky location ($\alpha_{l}$,~$\beta_{l}$)~of~$l^{th}$ of the BBH. $S$ is the area in two dimensions including all possible values of right ascension and declination. The number of BBHs is denoted as $N$.
%; i.e, in Cartesian coordinates $dS_m := dm^2)$.%
The density in excess $\delta_\mathrm{0}$ is written as
\begin{equation}
\delta_{\mathrm{0}}(\boldsymbol{m}):=P_{\mathrm{0}}(\boldsymbol{m}) / \bar{P}_{\mathrm{0}}-1=S P_{\mathrm{0}}(\boldsymbol{m})-1. 
\label{eq:delta_true}
\end{equation}
The true ACF is written as 
\begin{equation}
\omega_{0}(\boldsymbol{m}, \boldsymbol{n})=\left\langle\delta_{\mathrm{0}}(\boldsymbol{m}) \delta_{\mathrm{0}}(\boldsymbol{n})\right\rangle=S^{2}\left\langle P_{\mathrm{0}}(\boldsymbol{m}) P_{\mathrm{0}}(\boldsymbol{n})\right\rangle-1,
\label{10}
\end{equation}
where $\boldsymbol{m}$ and $\boldsymbol{n}$ are two independent points. $\bar{P}_{\mathrm{0}} = \frac{1}{S}$ is the area-averaged probability and $\left<~\right>$ denotes ensemble averages. By using Equation \ref{8}, $\left\langle P_{\mathrm{0}}(\boldsymbol{m}) P_{\mathrm{0}}(\boldsymbol{n})\right\rangle$ can be written as \begin{equation}
\left\langle P_{\mathrm{0}}(\boldsymbol{m}) P_{\mathrm{0}}(\boldsymbol{n})\right\rangle=\frac{1}{N^2}\left\langle\sum_{l, n} \delta^{(2)}\left(\boldsymbol{m}-\boldsymbol{m}_{l}\right) \delta^{(2)}\left(\boldsymbol{n}-\boldsymbol{n}_{n}\right)\right\rangle.
\label{11}
\end{equation}
We calculate the smeared ACF of BBHs mergers with location measurement errors. The individual GW source-location posterior is considered as a Gaussian
Distribution, which is given by 
\begin{eqnarray}
P_{l}\left(\mathbf{\boldsymbol{p}}-\boldsymbol{m}_{l}, \Delta \boldsymbol{m}_{l},d_{l}\right)=\frac{1}{\sqrt{(2 \pi)^{2}\left|\mathbf{C}_{l}\right|}} \exp \left[-\frac{1}{2}\left(\mathbf{\boldsymbol{p}}-\boldsymbol{m}_{l}-\Delta \boldsymbol{m}_{l}\right)^{T}\right.  \left.\times \mathrm{C}_{l}^{-1}\left(\mathbf{\boldsymbol{p}}-\boldsymbol{m}_{l}-\Delta \boldsymbol{m}_{l}\right)\right],
\label{12}
\end{eqnarray}
where $\boldsymbol{m}_{l}$ and $C_{l}$, respectively, denote the true sky location and the covariance matrix for $l^{th}$ GW source-location posterior. We assume that the posterior is proportional to the square of the distance from the GW source $d_{l}$, and the scatter $\Delta \boldsymbol{m}_{l}$ induced by the detector noise meets the Gaussian distribution of mean value zero and covariance matrix $C_{l}$. We first marginalize $P_{l}\left(\mathbf{\boldsymbol{p}}-\boldsymbol{m}_{l},~d_{l},~\Delta \boldsymbol{m}_{l}\right)$ over $\Delta \boldsymbol{m}_{l}$: 
\begin{eqnarray}
P_{l}\left(\mathbf{\boldsymbol{p}}-\boldsymbol{m}_{l},~d_{l}\right)=\int d V_{\Delta \boldsymbol{m}_{l}} P\left(\Delta \boldsymbol{m}_{l}\right) P_{l}\left(\mathbf{\boldsymbol{p}}-\boldsymbol{m}_{l},~\Delta \boldsymbol{m}_{l},~d_{l}\right).  
\label{13}
\end{eqnarray}
%where the resulting posterior $P_{l}\left(\mathbf{\boldsymbol{p}}-\boldsymbol{m}_{l},~d_{l}\right)$ is a Gaussian distribution with mean $\boldsymbol{m}_{l}$ and covariance matrix $2C_{l}$.% 
Then we marginalize $P_{l}\left(\mathbf{\boldsymbol{p}}-\boldsymbol{m}_{l},~d_{l}\right)$ over $d_{l}$:
\begin{eqnarray}
P_{l}\left(\mathbf{\boldsymbol{p}}-\boldsymbol{m}_{l}\right)=\int d V_{d_{l}} P\left(d_{l}\right) P_{l}\left(\mathbf{\boldsymbol{p}}-\boldsymbol{m}_{l},~d_{l}\right).
\label{14}
\end{eqnarray}
\begin{equation}
  P\left(d_{l}\right) \equiv \left\lbrace \begin{array}{ll}~0 ~~~~~~0\leq d_{l} <r_{1}\\
                                              Ad_{l}^{2} ~~~~~~r_{1}\leq d_{l} < r_{2} \\
                                              ~0 ~~~~~~d_{l}\geq r_{2},
                             \end{array} \right.
\label{eq:Harr}
\end{equation}
where $r_{1}$ and $r_{2}$ are radii of a survey spherical shell area and $A$ is the normalization factor. The averages in Equation \ref{13} and \ref{14} can be calculated for the posterior (as opposed to the final ACF) only when the noise-induced shifts $\Delta \boldsymbol{m}_{l}$ are uncorrelated with the two dimensional locations $\boldsymbol{m}_{l}$. $\boldsymbol{m}_{l}$ can also be denoted as ($\alpha_{l}$,~$\beta_{l}$). $P_{l}\left(\mathbf{\boldsymbol{p}}-\boldsymbol{m}_{l}\right)$ is written as 
\begin{eqnarray}
\label{16}
P_{l}\left(\mathbf{\boldsymbol{p}}-\boldsymbol{m}_{l}\right)=\int d S_{\boldsymbol{m}} P_{l}(\mathbf{\boldsymbol{p}}-\boldsymbol{m}) \delta^{(2)}\left(\boldsymbol{m}-\boldsymbol{m}_{l}\right).
\end{eqnarray}
The total posterior of the sky location of the GW events is written as \begin{eqnarray}
\label{17}
P(\mathbf{\boldsymbol{p}})=\frac{1}{N} \sum_{l} P_{l}\left(\mathbf{\boldsymbol{p}}-\boldsymbol{m}_{l}\right).
\end{eqnarray}
In this probability field, the ACF is given by
\begin{eqnarray}
\langle P(\mathbf{\boldsymbol{p}}) P(\mathbf{\boldsymbol{q}})\rangle=\frac{1}{N^2} \left\langle\sum_{l, n} P_{l}\left(\mathbf{\boldsymbol{p}}-\boldsymbol{m}_{l}\right) P_{n}\left(\mathbf{\boldsymbol{q}}-\boldsymbol{n}_{n}\right)\right\rangle,
\label{18}
\end{eqnarray}
Where $\boldsymbol{p}$ and $\boldsymbol{q}$ are two independent points. By using Equation \ref{14} and \ref{16}, Equation \ref{18} is given by
\begin{equation}
\langle P(\mathbf{\boldsymbol{p}}) P(\mathbf{\boldsymbol{q}})\rangle =\frac{1}{N^2}\left\langle\sum_{l n} \int d S_{\boldsymbol{m}} P_{l}(\mathbf{\boldsymbol{p}}-\boldsymbol{m}) \delta^{(2)}\left(\boldsymbol{m}-\boldsymbol{m}_{l}\right)\right.\\ \left.\times \int d S_{\boldsymbol{n}} P_{n}(\mathbf{\boldsymbol{q}}-\boldsymbol{n}) \delta^{(2)}\left(\boldsymbol{n}-\boldsymbol{n}_{n}\right)\right\rangle. 
\label{19}
\end{equation}
Then the three assumptions are made as follows:
\begin{enumerate}
\item The posterior is assumed to be uncorrelated with the true position of GW events, so we can write 
\begin{equation}
\begin{array}{l}
\left\langle P_{l}(\mathbf{\boldsymbol{p}}-\boldsymbol{m}) P_{n}(\mathbf{\boldsymbol{p}}-\boldsymbol{n}) \delta^{(2)}\left(\boldsymbol{m}-\boldsymbol{m}_{l}\right) \delta^{(2)}\left(\boldsymbol{n}-\boldsymbol{n}_{n}\right)\right\rangle \\ = \left\langle P_{l}(\mathbf{\boldsymbol{p}}-\boldsymbol{m}) P_{n}(\mathbf{\boldsymbol{p}}-\boldsymbol{n})\right\rangle\left\langle\delta^{(2)}\left(\boldsymbol{m}-\boldsymbol{m}_{l}\right) \delta^{(2)}\left(\boldsymbol{n}-\boldsymbol{n}_{n}\right)\right\rangle.
\end{array}
\end{equation}
\item 
Since $P_{l}$ and $P_{n}$ are posterior probability distributions estimated from two independent GW events:
\begin{equation}
\left\langle P_{l}(\mathbf{\boldsymbol{p}}-\boldsymbol{m}) P_{n}(\mathbf{\boldsymbol{q}}-\boldsymbol{n})\right\rangle=\left\langle P_{l}(\mathbf{\boldsymbol{p}}-\boldsymbol{m})\right\rangle\left\langle P_{n}(\mathbf{\boldsymbol{q}}-\boldsymbol{n})\right\rangle.
\end{equation}
\item
The homogeneity of space is assumed (but not the isotropy of space when calculating ACF),
\begin{equation}
\left\langle P_{l}(\mathbf{\boldsymbol{p}}-\boldsymbol{m})\right\rangle=P(\mathbf{\boldsymbol{p}}-\boldsymbol{m}) \text { and }\left\langle P_{n}(\mathbf{\boldsymbol{q}}-\boldsymbol{n})\right\rangle=P(\mathbf{\boldsymbol{q}}-\boldsymbol{n}).
\end{equation}
\end{enumerate}
By using the assumptions above and Equation \ref{11}, we can deduce
\begin{equation}
\begin{array}{l}\langle P(\mathbf{\boldsymbol{p}}) P(\mathbf{\boldsymbol{q}})\rangle=\frac{1}{N^2} \int d S_{\boldsymbol{m}} \int d S_{\boldsymbol{n}} P(\mathbf{\boldsymbol{p}}-\boldsymbol{m}) P(\mathbf{\boldsymbol{q}}-\boldsymbol{n})\times\left\langle\sum_{l,n} \delta^{(2)}\left(\boldsymbol{m}-\boldsymbol{m}_{l}\right) \delta^{(2)}\left(\boldsymbol{n}-\boldsymbol{n}_{n}\right)\right\rangle\\~~~~~~~~~~~~~~~~~~~~~=\int d S_{\boldsymbol{m}} \int d S_{\boldsymbol{n}} P(\mathbf{\boldsymbol{p}}-\boldsymbol{m}) P(\mathbf{\boldsymbol{q}}-\boldsymbol{n})\left\langle P_{\mathrm{0}}(\boldsymbol{m}) P_{\mathrm{0}}(\boldsymbol{n})\right\rangle.\end{array}
\end{equation}
The density contrast in the field of probability is given by 
\begin{equation}
\delta'(\mathbf{\boldsymbol{p}})=P(\mathbf{\boldsymbol{p}}) / \bar{P}-1.
\end{equation}
By using Equation \ref{10} and \ref{18}, the smeared ACF in this field is written as 
\begin{equation}
\omega(\boldsymbol{p}, \boldsymbol{q})=\left\langle\delta'(\mathbf{\boldsymbol{p}}) \delta'(\mathbf{\boldsymbol{q}})\right\rangle=S^{2}\langle P(\mathbf{\boldsymbol{p}}) P(\mathbf{\boldsymbol{q}})\rangle-1=\int_{S} d S_{\boldsymbol{m}} \int_{S} d S_{\boldsymbol{n}} P(\mathbf{\boldsymbol{p}}-\boldsymbol{m}) P(\mathbf{\boldsymbol{q}}-\boldsymbol{n}) \omega_{\mathrm{0}}(\boldsymbol{m}, \boldsymbol{n}),
\end{equation}
where $\boldsymbol{p}, \boldsymbol{q}, \boldsymbol{m}, and ~\boldsymbol{n}$ are vectors of two-dimensional coordinate ($\alpha$, $\beta$) space.

We can calculate the smeared ACF $\omega(\boldsymbol{p}, \boldsymbol{q})$ directly by $\omega_{0}(\boldsymbol{m}, \boldsymbol{n})$. $\omega_{0}(\boldsymbol{m}, \boldsymbol{n})$ only depends on $\omega_{BBH}(\theta)$ due to the homogeneity of space. In a two-dimensional coordinate area $S$, $\boldsymbol{m}$, and $\boldsymbol{n}$ can be, respectively, expressed as ($\alpha_{1},~\beta_{1}$), ($\alpha_{2},~\beta_{2}$), and we can deduce
\begin{equation}
\omega(\boldsymbol{p}, \boldsymbol{q})=\left\langle\delta'(\mathbf{\boldsymbol{p}}) \delta'(\mathbf{\boldsymbol{q}})\right\rangle=S^{2}\langle P(\mathbf{\boldsymbol{p}}) P(\mathbf{\boldsymbol{q}})\rangle-1=\int_{S} d S_{\boldsymbol{n}} \int_{S} d S_{\boldsymbol{m}} P(\mathbf{\boldsymbol{p}}-\boldsymbol{m}) P(\mathbf{\boldsymbol{q}}-\boldsymbol{n}) \omega_{\mathrm{0}}(\theta).
\label{26}
\end{equation}
\begin{equation}
\begin{array}{l}\cos \theta=\sin \left(\frac{\pi}{2}-\operatorname{\beta}_{1}\right) \cos \left(\alpha_{1}\right) \sin \left(\frac{\pi}{2}-\operatorname{\beta}_{2}\right) \cos \left(\alpha_{2}\right)+ \\~~~~~~~~~~~~~ \sin \left(\frac{\pi}{2}-\operatorname{\beta}_{1}\right) \sin \left(\alpha_{1}\right) \sin \left(\frac{\pi}{2}-\operatorname{\beta}_{2}\right) \sin \left(\alpha_{2}\right)+\cos \left(\frac{\pi}{2}-\operatorname{\beta}_{1}\right) \cos \left(\frac{\pi}{2}-\operatorname{\beta}_{2}\right),\end{array}
\end{equation}
where $\omega_{\mathrm{0}}(\theta)=b^{2}\omega_{\mathrm{DM}}(\theta)$. $\omega_{\mathrm{DM}}(\theta)$ is a dark matter ACF, which can be calculated by Equation \ref{1}.
We transformed Equation \ref{26} and defined two new variables $\boldsymbol{a}=\boldsymbol{p}-\boldsymbol{m}$
and $\boldsymbol{b}=\boldsymbol{q}-\boldsymbol{n}$ :
\begin{equation}
\omega_{sm}(\boldsymbol{p},~\boldsymbol{q})=\int_{S} d S_{\boldsymbol{a}} \int_{S} d S_{\boldsymbol{b}} P(\boldsymbol{a}) P(\boldsymbol{b}) \omega_{0}(\theta),
\end{equation}
where 
%$\cos\theta=\sin [\frac{\pi}{2}-(x-a)_{\beta}] \cos [(x-a)_{\alpha}] \sin [\frac{\pi}{2}-(y-b)_{\beta}] \cos [(y-b)_{\alpha}]+ \sin [\frac{\pi}{2}-(x-a)_{\beta}] \sin [(x-a)_{\alpha}] \sin [\frac{\pi}{2}-(y-b)_{\beta}] \sin [(y-b)_{\alpha}]+ \cos [\frac{\pi}{2}-(x-a)_{\beta}] \cos [\frac{\pi}{2}-(y-b)_{\beta}]$. 
$\omega_{sm}(\boldsymbol{p},~\boldsymbol{q})$ is averaged over two-dimensional coordinate area $S$ for a given $\theta$ to get $\omega(\theta)$.

\begin{figure}
   \centering
   \includegraphics[width=\textwidth, angle=0]{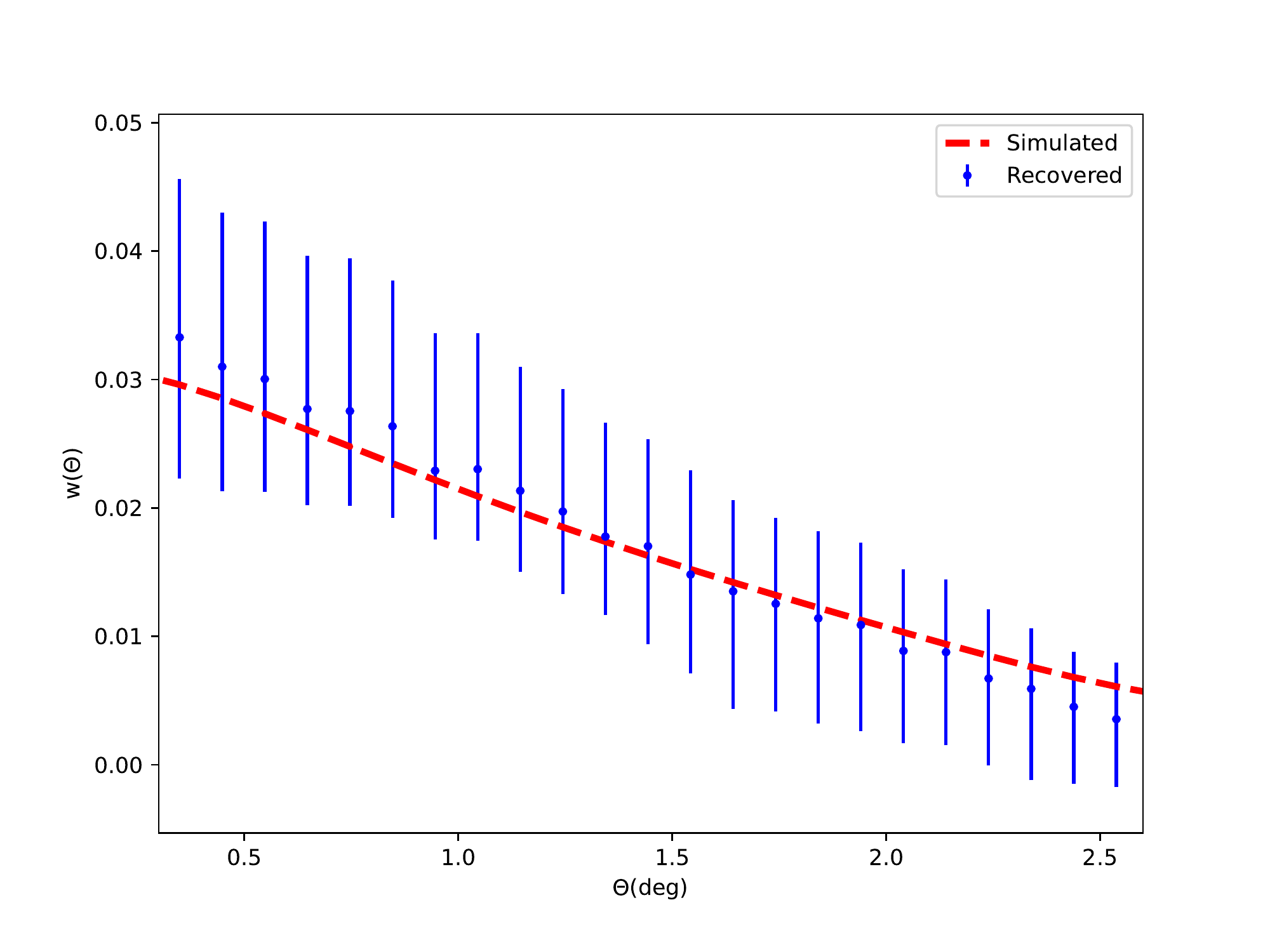}
   \caption{Using 5000 GW events in a $350~h^{-1}$ Mpc thick spherical shell simulated by setting the redshift to 0.3 and setting bias to 1.5.  The smeared theoretical ACF (red dashed line), which is scaled with input bias, is contrasted to the ACF calculated by the LS estimator (blue points with error bars). }
\label{sim}   
\end{figure}
   
\begin{figure}
\centering
\includegraphics[width=\textwidth, angle=0]{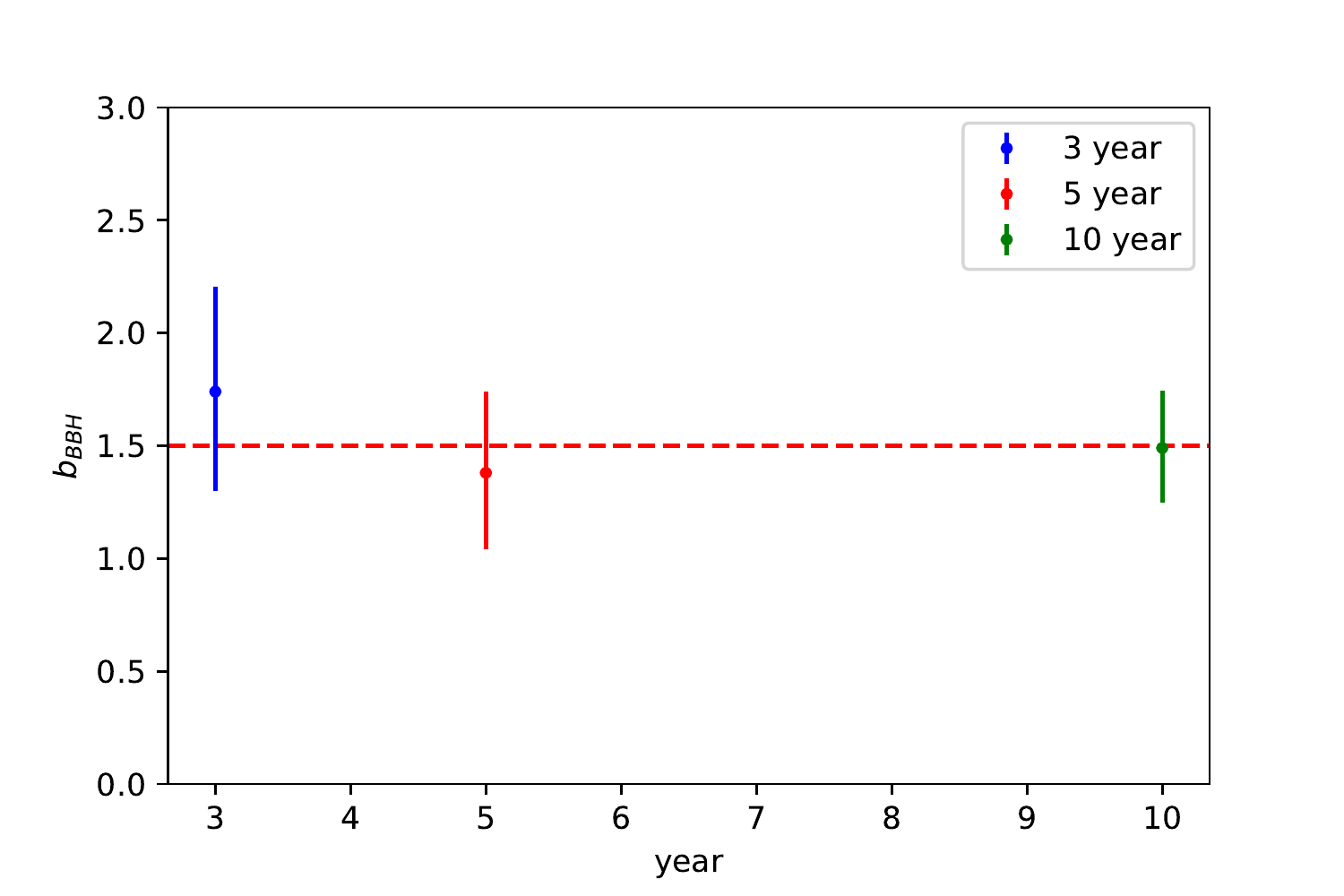}
\caption{By using the BBHs catalogs that were created with input bias factor 1.5 (red dashed line), the bias factor was recovered in a $350~h^{-1}$ Mpc thick spherical shell, at redshifts 0.3 using GW observations of BBHs in 3, 5, and 10 years. }
\label{f5}
\end{figure}

\section{Implementation method}
\label{im}
\subsection{The reference $C_0$}
\label{c0}
 { For simplicity, we assume all the GW event covariance matrices are the same, and consider a covariance matrix of sky location posteriors of GW150914 as $C_{0}$, \begin{eqnarray}
C_0= C_{GW150914}= \left(
\begin{array}{cccc}
985.7 & & -30.4 \\
-30.4 & & 46.5 
\end{array}     \right), 
\end{eqnarray}
where $C_{GW150914}$ is the covariance matrix of GW150914 event in the directions of right ascension and declination estimated by Bilby (\citealt{bilby}) using the data of Bilby's posterior samples for GW150914 (\citealt{150914}), which is observed by the LIGO Hanford (H1) and Livingston (L1) detectors.

Based on the discussion in \citet{23}, the requirement of location errors in ascension and declination should be within a degree and $a$ is scaled by the sensitivity improvements of detectors. The SNR for the advanced GW detector in Livingston and Hanford Observatory (L-H) during the first observing run (O1) is written as
\begin{equation}
\begin{aligned} \rho_{L-H}^{2}&=\rho_{1}^{2}+\rho_{2}^{2} \\ &=2\rho_{1}^{2}, 
\end{aligned}
\label{e1}
\end{equation}
where we assume $\rho_{1}^{2}=\rho_{2}^{2}$.
For ET-2CE (Cosmic Explorer locations: one in Hanford, USA and one in LIGO-India location. Einstein Telescope location: proposed one in Europe), the noise of the detector is 30 times less than that of L-H during the first observing run (\citealt{r12}; \citealt{r14}; \citealt{r15}), the SNR is written as
\begin{equation}
\begin{aligned} \rho_{ET-2CE}^{2}&=900\rho_{1}^{2}+900\rho_{2}^{2} +900\rho_{3}^{2} \\ &=2700\rho_{1}^{2}, 
\end{aligned}
\label{e2}
\end{equation}
where $\rho_{1}$, $\rho_{2}$, and $\rho_{3}$ are the SNR of the detectors, and we also assume $\rho_{1}^{2}=\rho_{2}^{2}=\rho_{3}^{3}$. By using Equation \ref{e1} and \ref{e2}, we can derive:
\begin{equation}
\frac{\rho_{ET-2CE}^{2}}{\rho_{L-H}^2} \approx 1500.
\label{e3}
\end{equation}
By using Equation \ref{rou}, the error in location for ET-2CE is written as, 
\begin{equation}
\Delta \Omega_{ET-2CE} =\frac{1}{1500} \Delta \Omega_{L-H}.
%=\frac{1}{1500} d^{2}\Delta \Omega_{0}.%
\end{equation} }
Thus, we set $a$=1/1500 for ET-2CE comparing with L-H during the first observing run. Then we calculate the black hole ACF smeared by the source-location posteriors from the perspective of theory and simulation.

\subsection{ Generation of BBHs catalogs} \label{subsec:sim}
{ To put our method to test, the code \textsc{lognormal\_galaxies} (\citealt{41})
%\cite{Agrawal:2017khv}%
is used to generate all-sky mock BBHs catalogs, which are produced for the log-normal distribution of the density field. With the BBHs bias factor used as input (\citet{pow}), the simulations of BBHs are obtained using the dark matter power spectrum given by \citet{36} and the Planck-18 cosmological parameters (\citealt{19b}). The bias factor is set to 1.5, and the redshift is set to 0.3. In real observation, comparing the value of the bias factor for BBHs and galaxies can reveal the environments of the BBH mergers. If $b_{BBH}=b_{gal}$, it means the BBH mergers happen in galaxies. If $b_{BBH}{\neq}b_{gal}$, it means BBHs are predominantly distributed outside galaxies, which can provide information about the formation mechanism of BBHs and their origin (\citet{23}). Further studies 
point out that if BBH mergers trace the distribution of dark matter more closely than the distribution of galaxies, it may imply that BBHs are of primordial origin; whereas, if BBH mergers trace the distribution of stars in luminous and massive galaxies more closely than other distribution, it may imply that BBHs are of stellar origin (\citet{bias}). Thus, the BBH catalogs could be generated and the ability 
to recover the bias could be shown. For simplicity, our conclusions are based on the fact that $b_{BBH}$ is independent of redshift, which remains valid when bias is redshift-dependent. The steps are listed below as follows: }
\begin{enumerate}
\item For the purpose that there are sufficient events, and the ACF does not
change significantly, a $350~h^{-1}$ Mpc thick spherical shell is chosen.
%,0.5,0.7,0.9]$%
%,0.14,0.16,0.18]$.%
$N$ BBHs are selected from this shell as the actual location of each GW event. Moreover, the posterior distributions of the GW events assumed Gaussian posteriors with the error matrix $C_{l}$. { The extent $\Delta z$ of the redshift bin corresponding to this shell thickness at redshift z = 0.3 turns out to be 0.13.}
%(see Section \ref{subsec:distance} for details regarding $C_{l}$). %
\item
\label{step2} To accurately simulate the locations of BBHs, one point is selected from each of the $N$ posteriors. Then, the ACF is estimated by the LS estimator. The process is repeated 10,000 times. $\omega_{BBH}$ is calculated by taking the average. 
\item 1000 BBH catalogs are generated to estimate the variance. First, 50 BBH catalogs are generated as realizations of the matter field, which give reasons for cosmic matter variance. From each of the 50 BBH catalogs, we randomly sample $N$ points as a sub-catalog and repeat this process 20 times so that we can get 20 sub-catalogs, which may account for the fluctuations due to the sample. Thus, there are 1000 BBH sub-catalogs. $\omega_{BBH}$ is estimated using the steps described above, that is, using one sub-catalog as the simulation of the actual GW events. Other sub-catalogs are used to estimate error bars on $\omega_{BBH}$ by the scatter of each $\omega_{BBH}$ calculated from 999 sub-catalogs. 
\item By comparing the recovered BBH ACF $\omega_{BBH}$ (which is calculated using the LS estimator) with the smeared theoretical BBH ACF (see Section \ref{subsec:thero}) using the least square fitting, the bias factor $b_{BBH}$ can be estimated. 
\end{enumerate}

{ On the one hand, when the number of GW events is inadequate, it does not reach the statistical effect, and our method is invalidated. On the other hand, when we calculate using the LS estimator, it only needs two-dimensional position coordinates for an angular correlation function, while it needs three-dimensional position coordinates for a two-point correlation function, which means that the angular correlation function does not fully utilize the position information of GW events as a two-point correlation function. The result in \citet[Figure 4]{23} shows that the bias factor can be recovered and the two-point correlation function can achieve an appropriate statistical effect using 5000 GW events. Therefore, we also tested whether the angular correlation function could also achieve an appropriate statistical effect using 5000 GW events to test the feasibility of our method and made a rough comparison between our method and that of \citet{23}.} Figure~\ref{sim} shows the smeared ACF contrasted to the ACF calculated by the LS estimator from a simulation using 5000 all-sky GW events in a $350~h^{-1}$ Mpc thick spherical shell, which is simulated by setting the redshift to 0.3 and the bias to 1.5. We found that our method is as feasible as the method of \citet{23} using 5000 GW events.

\section{Results and discussion} 
\label{sec:sum}
{ 
In Section \ref{sim}, we mentioned that errors in ascension and declination must be within a degree to ensure the validity of this method. We checked whether this demand could be met using ET-2CE detectors. For simplicity, we used the results of \citet[Figure 3]{23}, which showed the number of expected BBH mergers at various redshifts for a year of observations along with the fraction of events that are expected to be localized adequately for this type of study. The BBHs all-sky catalogs are generated using the same method in Section \ref{subsec:sim}. The observing time solely decides the number of GW events localized adequately for ET-2CE detectors with an observational time of 3,5 and 10 years at redshift z = 0.3. We quantitatively compared our results with the results of \citet{23}. Figure~\ref{f5} shows the ability to recover the bias factor using GW observations of BBHs in 3,5, and 10 years. The recovery of $b_{BBH}$ is coincidental with the bias within the error-bars. The bias can be recovered to within $\sim$ 52 $\%$ with an observational time of three years at redshift 0.3. For an observational time of five years, the bias can be recovered to within $\sim$ 51 $\%$. We can recover the bias factor to within $\sim$ 33 $\%$ with an observational time of ten years. With the increase in the observational time, the number of GW events increases, and the error bar further decreases. The errors in the bias recovery using our method are larger than that using the method of \citet{23}. On the one hand, it might result from the angular correlation, which does not fully utilize the position information of GW events. On the other hand, it might result from the different assumptions of the GW source-location posteriors and different values of the error matrices of GW events. The specific reasons need to be further studied in the future.

To explore the ability to probe large-scale structures using BBH observational data, we recover the bias factor by the ACF. First, we showed the feasibility of our method using 5000 BBH events and quantitatively analyzed the recovery effect of the bias factor using GW observations of BBHs in 3,5, and 10 years by the third-generation GW detectors and compared it with that of \citet{23}, which reveals the potential of GW detection as the probe in large scale structure in the future. It reveals the environment of the BBH mergers. Although the error of the estimated bias is still larger than that of the galaxy bias due to the large location error of GW events, using GW detections to construct large-scale structures is an important method to probe dark matter distribution. In future research, to expand our work, we can completely abandon the hypothesis that the position posterior distribution of GW events is a Gaussian Distribution. In addition, by using the real position posterior distribution of GW events to calculate the angular correlation function and the two-point correlation function, we can compare their ability to probe large-scale structures.}

\begin{acknowledgements}
%I would like to express my gratitude to all those who helped me during the writing of this thesis.
X. Shao is grateful to A. Vijaykumar for the statistical help and thanks  ZhengXiang Li and Huan Zhou for their contribution to the bias factor. We are very grateful to the anonymous referee for her/his valuable comments and suggestions. This work was supported by the National Natural Science Foundation of China under grant No.11922303.
X. Fan is supported by Hubei province Natural Science Fund for the Distinguished Young Scholars. 
\end{acknowledgements}

\bibliographystyle{raa}
\bibliography{ms2021-0212.bbl}

\end{document}